\documentclass[aps,pre,floatfix,nofootinbib,preprint
]{revtex4}
\usepackage{graphicx} \usepackage{amsmath} \usepackage{amssymb}
\usepackage[utf8]{inputenc}
\usepackage{xcolor}
\newcommand{\be}{\begin{eqnarray}}
\newcommand{\ee}{\end{eqnarray}}

\newcommand{\I}{{\rm I}}
\newcommand{\II}{{\rm II}}
\newcommand{\III}{{\rm III}}
\newcommand{\IV}{{\rm IV}}
\newcommand{\comment}[1]{}
\newcommand{\A}{{\cal A}}

\begin{document}

\title{Regret theory, Allais' Paradox, and Savage's omelet}

\author{ V.G. Bardakhchyan$^{1,2)}$ and A.E. Allahverdyan$^{1,2)}$ }

\address{ 
$^{1)}$Alikhanian National Laboratory (Yerevan Physics Institute), Alikhanian Brothers Street 2,  Yerevan 0036, Armenia,\\
$^{2)}$Yerevan State University, 1 A. Manoogian street, Yerevan 0025, Armenia
}

\date{\today}

\begin{abstract}
We study a sufficiently general regret criterion for choosing between two probabilistic lotteries. For independent lotteries, the criterion is consistent with stochastic dominance and can be made transitive by a unique choice of the regret function. Together with additional (and intuitively meaningful) super-additivity property, the regret criterion resolves the Allais' paradox including the cases were the paradox disappears, and the choices agree with the expected utility. This super-additivity property is also employed for establishing consistency between regret and stochastic dominance for dependent lotteries. Furthermore, we demonstrate how the regret criterion can be used in Savage's omelet, a classical decision problem in which the lottery outcomes are not fully resolved. The expected utility cannot be used in such situations, as it discards important aspects of lotteries.

{\bf Keywords:} Regret theory, Allais' paradox, stochastic dominance, transitive regret.

{\bf JEL Classification:} D81.

\end{abstract}

\maketitle

\section{Introduction}

The history of expected utility theory (EUT) started with Bernoulli's work resolving the St. Petersburg paradox \cite{stearns}. Several axiomatic schemes for EUT are known \cite{luce,savage}. Currently, EUT has applications in a wide range of fields, including economics \cite{Hey_theories}, psychology \cite{baron}, evolutionary game theory \cite{sigmund}, and general artificial intelligence \cite{everitt}.

EUT shows how to choose between two lotteries \cite{luce,savage,Hey_theories}:
\begin{gather}
\label{1}
(x,p)=
\begin{pmatrix}
x_{1} & x_{2} & ... & x_{n}\\
p_{1} & p_{2} & ... & p_{n}
\end{pmatrix},\qquad
(y,q)=
\begin{pmatrix}
y_{1} & y_{2} & ... & y_{n}\\
q_{1} & q_{2} & ... & q_{n}
\end{pmatrix},\\
{\sum}_{k=1}^np_k={\sum}_{k=1}^nq_k=1,
\end{gather}
where $(p_1,...,p_n)$ and $(q_1,...,q_n)$ are (resp.) the probabilities of monetary outcomes
$(x_1,...,x_n)$ and $(y_1,...,y_n)$ within each lottery. EUT proposes the following functional for each lottery \cite{luce,savage,Hey_theories}:
\begin{equation}
\label{2}
V(x, p) = {\sum}_{i=1}^{n}u(x_{i})p_{i}, 
\end{equation}
where $u(x_i)$ is the utility of the monetary value $x_i$. EUT recommends choosing in (\ref{1}) the first lottery, if $V(x,p)>V(y,q)$. \comment{This recommendation holds even for a single choice between $(x,p)$ and $(y,p)$, though intuitively $V(x,p)$ corresponds to the average utility gained, when the lottery $(x,p)$ is chosen many times.} 

Experiments revealed problems with EUT and its axiomatic foundations. In particular, several classic experiments cannot be explained by EU for any choice of the utility function $u(.)$ in (\ref{2}) \cite{allais,Machina}. People generally choose in contradiction to EUT, violating the independence axiom, one of four axioms of the von Neumann-Morgenstern formulation of EUT \cite{luce}. The most prominent example of this is Allais's paradox \cite{allais}, where each human subject chooses between two lotteries. The prospect theory \cite{Prospect,Cumul_Prospect}, and rank-dependent utility theory \cite{Quiggin,wak} discarded the independence axiom, and proposed functionals similar to $V(x, p)$ in (\ref{2}), where instead of probabilities $p_i$ one employs weights $\pi_i$ that generally depend both on $(p_1,...,p_n)$ and $(x_1,...,x_n)$. Refs.~\cite{Machina,Hey_theories,baron} discuss these and other alternatives to EUT. 

There are also other situations where EUT does not apply. EUT cannot be used directly when the lottery outcome remains uncertain even after the lottery choice has been made. A good example of this situation is the decision problem known as Savage's omelet \cite{savage}. In our knowledge, this problem has never been studied from a viewpoint of EUT's inapplicability.

As we show below, both Allais' paradox and Savage's omelet can be resolved by the regret theory (RT), which is one of the alternatives of EUT. The main difference of RT compared to EUT is that RT does not operate with a value functional for a single lottery. Instead it counter-factually compares two lotteries. RT has an intuitive emotional appeal, and it is also related to cognitive aspects of decision making \cite{bourgeois2010regret}. RT was first proposed by Savage in minimax form \cite{savage} [see \cite{tempered} for an update of this approach], and later brought to its current form in \cite{Regret_Sugden,Bell}; see \cite{Regret_main,Machina} for a review.
Ref.~\cite{Regret_Sugden} extended the regret to independent lotteries and noted its potential in explaining Allais' paradox. Ref.~\cite{Regret_Sugden} also analyzed transitivity, common ration effect, and preference reversals. Functional forms involving two lotteries were given axiomatic foundation in \cite{Fishburn_SSB}. An axiomatic formulation of regret was attempted in \cite{Reg_axioms}. 

This work has three purposes. First, we want to show how Allais' paradox is solved by a transitive and super-additive RT. People mentioned regret in the context of Allais' paradox [see e.g. \cite{baron,Regret_Sugden,bourgeois2010regret}], but so far no systematic and complete solution of this paradox was provided. Our solution is rather complete, because it also predicts conditions under which the paradox does not hold. Both transitivity and super-additivity have transparent meaning for regret theories in general. We do clarify their applicability range. This is especially important for transitivity, because generally regret theories do not lead to transitive predictions \cite{Starmer}. 

Second, we prove that the transitive and super-additive regret theory is consistent with the stochastic dominance criterion \cite{luce}. Stochastic dominance is a useful tool, but it does not apply to comparing any pair of lotteries. The previous literature in this direction is mostly negative showing that regret-based approaches violate first order stochastic dominance \cite{Quiggin_std, levy2017regret} \footnote{Ref.~\cite{Quiggin_std} analyzed relations between RT and stochastic dominance for a specific case. This analysis is based on more general formulation of first order stochastic dominance that compares cumulative distribution functions. Here we focus on the simplest version of stochastic dominance.}. Third, we demonstrate|using as an example Savage's omlet problem|that RT can recommend choosing between lotteries with not resolved outcomes, a task which cannot be consistently addressed by EUT. 

The paper is organized as follows. Section \ref{regret_basic} is devoted to regret functional for independent lotteries and some of its properties related to the expected utility. In decision making theory the functional form is frequently derived from axiomatic foundation. In contrast, here we first introduce the functional considered, then derive its properties. Section \ref{alais_solution} is devoted to Allais' paradox and its relations to other concepts. Stochastic dominance abidance is considered in section \ref{regret_sto}. Section \ref{ap_a} analyzes Savage's omelet problem, identifies an aspect that prevents the applicability of the expected utility theory, and solves this problem via the regret. We summarize in the last section.

\comment{
In most economic literature whenever the outcomes are numbers (from $R$), lottery is called simple. 
Outcomes themselves can be lotteries, and in that case, lottery is called compound.
However, and compound lottery can ultimately be given in simple form.
So we concentrate on latter ones.
In that form each lottery is discrete random variable with finite values.}

\section{Regret and its features}
\label{regret_basic}

\subsection{Axioms of Expected Utility Theory (EUT)}

We remind the four axioms of EUT (\ref{2})|completeness, transitivity, continuity, independence|since they will motivate our further consideration. First of all one introduces a preference relation $\succeq$, and indifference relation $\sim$ between the lotteries (\ref{1}), where $\sim$ means that both $\succeq$ and $\preceq$ hold. When comparing two lotteries in (\ref{1}) we sometimes assume (without loss of generality) the same outcomes: $\{x_k=y_k\}_{k=1}^n$. If they are initially different, we can introduce suitable zero-probability events and make them identical.

{\bf 1.} The completeness axiom states that any pair of lotteries in (\ref{1}) can be compared:
\be
(x,p) \succeq (x,q)\quad {\rm or}\quad (x,q) \succeq (x,p)
\quad {\rm or}\quad (x,q) \sim (x,p),
\label{comp}
\ee
where $(x,p) \succeq (x,q)$ means that lottery $(x,q)$ is not preferred to $(x,p)$.

{\bf 2.} The transitivity axiom states:
\be
\label{trans}
(x,p) \succeq (x,q)\succeq (x,r)\quad {\rm means}\quad (x,p) \succeq (x,r),
\ee

{\bf 3.} The continuity axiom states for any three lotteries
\begin{gather}
(x,p) \succeq (x,q)\succeq (x,r)\quad {\rm implies}\quad (x,q)\sim (x,
\alpha p+(1-\alpha)r), 
\label{cont}
\end{gather}
for some $\alpha \in [0,1]$. 
This axiom implies continuity of the value function to be deduced from the four axioms.

{\bf 4.}
The independence axiom|also known independence of irrelevant alternatives or the sure-thing principle|claims that combining each of two lotteries with any fixed one will not alter the preferences \cite{baron,jeffrey}: 
\be
\label{inde}
(x,p) \succeq (x,q) \quad {\rm means}\quad (x,\alpha p+(1-\alpha)r ) \succeq (x,\alpha q+(1-\alpha)r),
\ee
where the irrelevant alternative is $(x,r)$. 
Eq.~(\ref{inde}) is among the most controversial axioms in decision theory and has triggered many debates \cite{baron,jeffrey}; see in this context also Appendix \ref{inde3}, where we explain why specifically the meaning of (\ref{inde}) can be ambiguous. Ref.~\cite{pearl} briefly reviews its current status with counter-examples. Experimental studies showed violations of (\ref{inde}), with some concerns on whether these violations are systematic \cite{baron}. 

\subsection{Definition of regret}

The regret defines a counterfactual outcome-wise comparison between the lotteries (\ref{1}) using certain ideas of EUT. Hence for particular cases it would coincide with the decision criterion of EUT. The utility function $u(x)$ is assumed to exist beforehand and known to the decision-maker \cite{baron}.

Assume that $(y,q)$ is chosen and its outcome $y_j$ is found. The decision-maker compares this outcome with what would be found if $(x,p)$ would be taken and defines:
\be
R(x,p;y_j)\equiv{\sum}_{i=1}^n f(u(x_i)-u(y_j))p_i,
\label{r1}
\ee
where $u(x)$ is the utility function, and $f(x)$ is a function holding
\be
\label{sign}
f(x\geq 0)\geq 0, \qquad f(x\leq 0)\leq 0, \qquad f(0)=0.
\ee
In particular, $R(x,p;y_j)>0$ (positive regret), if $x_i>y_j$. Generally, $f(x)$ accounts for both regret and appreciation. We get a pure regret (appreciation), if $f(x\leq 0)=0$ ($f(x\geq 0)=0$).

Since $(x,p)$ was not actually chosen, its outcomes are not known; hence the averaging in (\ref{r1}). Moreover, once the decision-maker keeps on choosing $(y,q)$ and explores all its outcomes according to their probabilities, the average of (\ref{r1}) reads:
\be
R(x,p;y,q)\equiv{\sum}_{j=1}^nq_j R(x,p;y_j)={\sum}_{i,j=1}^n f(u(x_i)-u(y_j))p_iq_j,
\label{r3}
\ee
where (\ref{r3}) already assumed that the events $(y_j,x_i)$ are independent, i.e. their joint probability is $q_jp_i$. This additional information is to be provided for unambiguous definition of lotteries in (\ref{1}). 

Note that (\ref{r1}, \ref{r3}) are asymmetric with respect to the lotteries (\ref{1}), because $(y,q)$ is actually chosen, while $(x,p)$ is reasoned counter-factually given this choice. The regret preference $\succeq_{\rm reg}$ is defined as \cite{Regret_Sugden,Bell,Regret_main,Machina}
\be
\label{r7}
(x,p)\succeq_{\rm reg}(y,q)\quad {\rm iff}\quad R(y,q;x,p)-R(x,p;y,q)=
{\sum}_{i,j=1}^n g(u(y_j)-u(x_i))p_iq_j\leq 0,
\ee
where 
\be
\label{r8}
g(x)\equiv f(x)-f(-x),
\ee
is anti-symmetric and monotonic:
\be
\label{u2}
&& g(x) = - g(-x),\\
&& g(x)\geq g(y)\quad {\rm for}\quad x\geq y.
\label{uu2}
\ee
The meaning of $R(y,q;x,p)-R(x,p;y,q)\leq 0$ is that $(x,p)$ is preferred if its leads to a smaller average regret.
For a particular case 
\be
g(x)=ax, \qquad a>0,
\label{rev}
\ee
where $a$ is a constant, we revert from (\ref{r7}) to the expected utility. Note that (\ref{rev}) is achieved for various functions $f(x)$; e.g. $f(x)=ax/2$ or $f(x)=a\,{\rm max}[x,0]$. 

The above definition generalizes for a non-trivial joint probability $P(x_i,y_j)$ of $(x_i,y_j)$ with 
\be
\label{voche}
{\sum }_{i=1}^nP(x_i,y_j)=q_j,\qquad 
{\sum }_{j=1}^n P(x_i,y_j)=p_i.
\ee
Now $p_i$ in (\ref{r1}) should be replaced by conditional probability $P(x_i|y_j)$, which is reasonable for a counter-factual reasoning, and instead of (\ref{r1}--\ref{r7}) we have
\be
\label{r4}
&&R(x,p;y_j)\equiv{\sum}_{i=1}^n f(u(x_i)-u(y_j))P(x_i|y_j),\\
\label{r44}
&&R(x,p;y,q)\equiv{\sum}_{j=1}^n q_j R(x,p;y_j)={\sum}_{i,j=1}^n f(u(x_i)-u(y_j))P(x_i,y_j),\\
&& (x,p)\succeq_{\rm reg}(y,q)\quad {\rm iff}\quad {\sum}_{i,j=1}^n g(u(y_j)-u(x_i))P(x_i,y_j)\leq 0.
\label{r777}
\ee
In particular, the outcomes in (\ref{voche}) can refer to the same states of nature \cite{Reg_axioms,luce,jeffrey}. This implies 
\be
P(x_i,y_j)=p_i\delta_{ij},\quad i,j=1,...,n,
\label{r55}
\ee
where $\delta_{ij}$ is the Kroenecker delta, and where $\{p_i=q_i\}_{i=1}^n$ are the probabilities for those unknown states of nature; see section \ref{ap_a} for details.

\comment{In most of the literature the regret function - $f$ is deemed to be increasing and concave. (Mostly no arguments are brought forward for concavity, as it has nothing to do with risk aversion.) \cite{Machina}.
Note that some authors take $r(x,y) = -g(u(x)-u(y))$ as the function of regret (see for example \cite{Machina}). Thus, their function is positive when first lottery is preferred to second one.}

\subsection{Two propositions about regret}

Note that for the regret preference relation (\ref{r7}) we can take lotteries (\ref{1}) to have the same outcomes, $x_k=y_k$, using the same argument as before (\ref{comp}). Now the completeness axiom (\ref{comp}) obviously holds for $\succeq_{\rm reg}$. The continuity axiom is valid as well. 

\vspace{0.1cm}
\underline{\textbf{Proposition 1.}} For the regret preference relation (\ref{r7})
\be
(x,p) \succeq_{\rm reg} (x,q)\succeq_{\rm reg} (x,r)\quad {\rm implies}\quad (x,q)\sim_{\rm reg} (x,\alpha p+(1-\alpha)r),
\label{orde}
\ee
for some $\alpha\in [0,1]$. Working out the last relation in (\ref{orde}) we find
\be
&& \alpha=B/(A+B) \in [0,1],\\
&& A={\sum}_{i,j=1}^n p_{i}q_{j}g(u(x_i)-u(x_j))\geq 0,\qquad 
B={\sum}_{i,j=1}^n r_{i}q_{j}g(u(x_j)-u(x_i))\geq 0,
\label{assum}
\ee
where (\ref{assum}) follows from first and second relations in (\ref{orde}).
\vspace{0.1cm}

It is known that $\succeq_{\rm reg}$ violates transitivity for a general choice of $f(x)$ \cite{Starmer}.
In particular, the transitivity is violated under (\ref{r55}) \cite{Transitive_regret}; e.g. for the same states of nature. Transitivity violation is not necessarily a drawback, since there are arguments for involving 
non-transitive choices even in normative choices \cite{Fishburn_nontransitive}. Ref.~\cite{Transitive_regret_ind} shows that for the most general form of regret there exist models not violating transitivity. 

Let us now provide a sufficiently complete solution for the transitivity of $\succeq_{\rm reg}$.
First, we show that $\succeq_{\rm reg}$ will be transitive for a particular choice of $f(x)$ in (\ref{r7}). Define 
\be
\label{chin0}
f(x)=b(a^x-1), 
\ee
where $a>0$ and $b>0$. Eq.~(\ref{sign}) holds. Now $(x,p)\succeq_{\rm reg}(x,q)$ amounts to 
\begin{gather}
\label{chin1}
v(p)w(q)\geq v(q)w(p),\\
v(p)\equiv{\sum}_{i=1}^n a^{u(x_i)}p_i>0,\qquad w(q)\equiv{\sum}_{i=1}^n a^{-u(x_i)}q_i>0.
\label{chin2}
\end{gather}
Eqs.~(\ref{chin1}, \ref{chin2}) imply that with the choice (\ref{chin0}), $\succeq_{\rm reg}$ is transitive. Fisburn's theorem on transitivity \cite{Fishburn} shows that (\ref{chin0}) is also necessary for transitivity.

\vspace{0.2cm}
\underline{\textbf{Proposition 2.}} The regret preference relation $\succeq_{\rm reg}$ given by (\ref{r7}) preserves transitivity iff (\ref{chin0}) holds.
\vspace{0.2cm}

Returning to (\ref{comp}--\ref{inde}) we see that only the independence axiom can be violated by $\succeq_{\rm reg}$; see below for more details.

\section{Solving Allais' paradox with regret}
\label{alais_solution}

There was a great deal of attention focused on Allais' paradox as one of the major systematic violations of  EUT \cite{baron,allais,Prospect,Cumul_Prospect,Machina,schoemaker}. Regret theory is mentioned in the context of Allais's paradox \cite{baron,bourgeois2010regret,Regret_Sugden}, but no systematic solution of the paradox via the regret theory was so far provided. We show below that this solution can be achieved by respecting the transitivity and that it does provide an important constraint on the form of $g(x)$ in (\ref{r7}, \ref{r8}).

Consider the standard formulation of the Allais' paradox \cite{baron,allais}. A decision make is choosing between the following two lotteries [cf.~(\ref{1})]:
\begin{equation}
\label{al_1}
\I\equiv \begin{pmatrix}
    1\\
    1
    \end{pmatrix},\qquad \II\equiv
    \begin{pmatrix}
    0 & 1 & 5\\
    0.01 & 0.89 & 0.1
    \end{pmatrix},
\end{equation}
and then between 
\begin{equation}
\label{al_2}
 \III\equiv   \begin{pmatrix}
    0 & 1 \\
    0.89 & 0.11
    \end{pmatrix} ,\qquad
    \IV\equiv     \begin{pmatrix}
    0 & 5\\
    0.9 & 0.1
    \end{pmatrix},
\end{equation}
where the monetary outcomes in (\ref{al_1}, \ref{al_2}) are normally given in millions of \$. 

There are 4 possible outcomes here: $(\I, \III)$, $(\I, \IV)$, $(\II, \III)$, $(\II, \IV)$, where $(\I, \III)$ means choosing $\I$ in (\ref{al_1}) and $\III$ in (\ref{al_2}). Choosing  $(\I, \III)$ or $(\II, \IV)$ is consistent with the EUT; e.g. $(\I, \III)$ is achieved if $u(1)<u(5)$ and $u(1)\approx u(5)$. In contrast, most of people take $(\I, \IV)$ thereby violating the expected utility theory (EUT) \cite{baron}. 

Applying preference relation (\ref{r7}) to the choice $(\I, \IV)$, we will find an important and intuitive condition for function $g(x)$. Now $\I\succeq_{\rm reg}\II$ reads from (\ref{r7}):
\begin{equation}
        0.01 \cdot g(u(0)-u(1))+0.1 \cdot g(u(5)-u(1)) < 0.
    \label{al_3}
\end{equation}
Since $g(x)$ is an increasing function [cf.~(\ref{uu2})], (\ref{al_3}) implies
\begin{equation}
    \label{al_4}
u(5)-u(1) < u(1) - u(0). 
\end{equation}
Thus (\ref{al_4})|which can be realized with a concave function $u(x)$ and hence relates to risk-aversion|is a necessary condition for (\ref{r7}) to explain Allais' paradox. Likewise, demanding ${\rm IV}\succeq_{\rm reg}{\rm III}$ in (\ref{al_2}) we get
\begin{equation}
    \label{al_5}
    0.089 \cdot g(u(5)-u(0)) - 0.099 \cdot g(u(1)-u(0)) + 0.011 \cdot g(u(5)-u(1))>0
\end{equation}

\noindent Taking the difference of (\ref{al_5}) and (\ref{al_3}) we get
\begin{equation*}
    -0.089 \cdot g(u(5)-u(0)) + 0.089 \cdot g(u(1)-u(0)) + 0.089 \cdot g(u(5)-u(1))<0,
\end{equation*}
yielding 
\begin{equation}
    \label{al_7}
    g(u(5)-u(0)) > g(u(1)-u(0)) + g(u(5)-u(1)).
\end{equation}
Now (\ref{al_7}) is the second necessary condition for solving Allais's paradox. Taking (\ref{al_7}) and (\ref{al_3}) together is necessary and sufficient for solving the paradox. It is intuitively clear what (\ref{al_7}) means. The decision maker is more impressed (i.e. experiences more regret) with the difference $u(5)-u(0)$, than with this difference $u(5)-u(0)=u(1)-u(0)+u(5)-u(1)$ coming in two separate pieces: $u(1)-u(0)$ and $u(5)-u(1)$. We rewrite (\ref{al_7}) as a more general condition:
\begin{equation}
    \label{al_8}
    g(x+y) \geq g(x) + g(y),\quad x\geq 0, \quad y\geq 0,
\end{equation}
which is the super-additivity (in positive domain) for $g(x)$. Noting from (\ref{u2})
that $g(0)=0$, we recall that any convex function $g(x)$ with $g(0)=0$ is super-additive \footnote{This fact should be known, but let us present its short proof. First note that $g(tx)\leq tg(x)$ for $0<t<1$ due to $g(t(x)+(1-t)\cdot 0)\leq tg(x)+(1-t)g(0)=tg(x)$. Next, $g(x)+g(y)=g\left((x+y)\frac{x}{x+y}\right)+ g\left((x+y)\frac{y}{x+y}\right)\leq \frac{x}{x+y} g(x+y)+\frac{y}{x+y}g(x+y)=g(x+y)$.}. A simple example of a function that is easily shown to be super-additive, but is not convex is $g(x)=x\,e^{-x^{-2}}$ \cite{japan}. Indeed, $\frac{{\rm d}^2}{{\rm d}x^2}g(x)=2\,e^{-x^{-2}}\,x^{-5}(2-x^2)$, i.e. $g(x)$ is concave (convex) for $x>\sqrt{2}$ ($\sqrt{2}>x>0$) 
\footnote{Ref.~\cite{Reg_axioms} mentioned the super-additivity condition in the context of regret. Ref.~\cite{Regret_Sugden} employed convexity (concavity) features of regret functional, but without any definite reason.}. We formulate our results as follows.

\vspace{0.2cm}
\underline{\textbf{Proposition 3.}} Allais's paradox can be explained by regret, if and only if function $g(x)$ in (\ref{r7}) is strongly super-additive for some values in positive domain.
\vspace{0.2cm}

\textbf{Example}. We take the transitive regret and logarithmic utility [cf.~(\ref{u2}, \ref{chin0})]
\be
\label{ex}
g(x)=\sinh \left(\frac{x}{\beta}\right),\qquad u(x)=\ln\left(\frac{x}{\gamma}+1\right),
\ee
where $\beta>0$ and $\gamma>0$ are positive parameters that characterize the decision maker. Here $\gamma>0$ defines the threshold of the concave (risk-averse) utility $u(x)$ ($u(0)=0$), because only for $\frac{x}{\gamma}\ll 1$ we have $u(x)\simeq 0$. In a sense, $\gamma$ defines the initial money, since only for $\frac{x}{\gamma}\gtrsim 1$ the decision maker will care about money. Likewise, $\beta$ has a similar meaning of threshold, but for the regret function: if $\frac{x}{\beta}\ll 1$, then $g(x)=\sinh (\frac{x}{\beta})\simeq \frac{x}{\beta}$ is effectively in the regime EUT. 

Now $g(x)$ in (\ref{ex}) holds super-additivity condition (\ref{al_8}), since $\sinh (0)=0$ and $\frac{{\rm d}^2}{{\rm d}x^2}\sinh(x)=\sinh(x)\geq 0$ for $x\geq 0$; hence (\ref{al_7}) holds. For solving Allais' paradox we need to look at
condition (\ref{al_3}), which from (\ref{ex}) amounts to 
\be
&& \gamma<\zeta(\beta), \\ 
&& \zeta(\beta\to\infty)=5^{-10},\quad \zeta(1)=0.021,  
\quad \zeta(\beta\to 0)=1/3.
\label{deb}
\ee
Hence $\zeta(\beta)$ changes from $5^{-10}$ to $1/3$, when $\beta$ moves from $\infty$ to $0$.
Let us focus on $\gamma<0.021$ in (\ref{deb}). We know that (\ref{al_1}, \ref{al_2}) are to be given in millions of \$. Hence we multiply both $x$ and $\gamma$ in $u(x)=\ln(\frac{x}{\gamma}+1)$ by $10^6$, and reach the following conclusion: starting from the initial money $\geq 21 000$ \$ the decision maker will behave according to the expected utility and choose lotteries $(\II, \IV)$ in (\ref{al_1}, \ref{al_2}). The interpretation of 
the other two values of $\zeta(\beta)$ in (\ref{deb}) is similar. Note in this context that $5^{-10}$ is equivalent to $5^{-10}\times 10^8\simeq 10$ cents.  

It is reported that with smaller outcomes|not millions of \$ in (\ref{al_1}, \ref{al_2})|Allais' paradox need not hold \cite{Small_outcomes, camerer1989experimental, blavatskyy2015now}. Other authors note that when shifting all outcomes in (\ref{al_1}, \ref{al_2})  with the same substantial positive amount, Allais' paradox will not hold (aversion of "0" outcome) \cite{0_aversion}. The scheme given by (\ref{ex}) handles both experimental results.

\textbf{Remark 1}.
The super-additivity (\ref{al_8}) of $g(x)$ (and its ensuing relations with convexity) does not relate to risk-aversion and risk-seeking, as defined via utility $u(x)$. To understand this, compare the following two lotteries:
\begin{equation}
    \begin{pmatrix}
    x \\
    1
    \end{pmatrix}     \quad {\rm and} \quad
    \begin{pmatrix}
    x-\epsilon & x+\epsilon \\
    0.5 & 0.5
    \end{pmatrix}, \qquad \epsilon>0.
\end{equation} 
Now the first (certain) lottery is regret-preferable compared with the second (uncertain) lottery if 
$g(u(x)-u(x-\epsilon))>-g(u(x)-u(x+\epsilon))$, which is achieved due to a monotonically increasing $g(x)$, and concavity of $u(x)$; i.e. the risk-aversion at the level of the utility. Likewise, the convexity of $u(x)$ (risk-seeking utility) will lead to preferring the uncertain lottery. 

{\bf Remark 2.} Note that the regret is invariant with respect to $u(x)\to u(x)+a$, where $a$ is arbitrary, but it is not invariant with respect to $u(x)\to b u(x)$, where $b>0$; see e.g. the very example (\ref{ex}). After transformation $u(x)\to b u(x)$, one can redefine $g_b(x) = g(bx)$ such that the regret stays invariant. This redefinition respects transitivity and super-additivity of $g(x)$. 

{\bf Remark 3.}
Recall that the independence axiom (\ref{inde}) (or the axiom of irrelevant alternatives) is the main axiom of EUT violated by the regret theory. Allais' paradox can be reformulated in such a way that the presence of this axiom is made obvious. To this end one writes (\ref{al_1}, \ref{al_2}) as
\begin{align}
    \label{re1}
&\I= \begin{pmatrix}
    1    & 1   & 1\\
    0.01 & 0.1 & 0.89
    \end{pmatrix}, &
\III= \begin{pmatrix}
    1    & 1   & 0\\
    0.01 & 0.1 & 0.89
    \end{pmatrix},\\
&\II= \begin{pmatrix}
    0    & 5   & 1\\
    0.01 & 0.1 & 0.89
    \end{pmatrix},  &
\IV= \begin{pmatrix}
    0    & 5   & 0\\
    0.01 & 0.1 & 0.89
    \end{pmatrix}.
    \label{re2}
\end{align}
We emphasize that $\I$ and $\II$ in (\ref{re1}, \ref{re2}) (as well as $\III$ and $\IV$) refer to independent events. 

It is seen that $\I$ and $\II$ have the common last column $({1\atop 0.89})$, while for $\III$ and $\IV$ the common last column is $({0\atop 0.89})$. These last columns (i.e. the corresponding outcomes with their probabilities) plays the role of independent alternatives. If they are deemed to be irrelevant, e.g. $({1\atop 0.89})$ is irrelevant when deciding between $\I$ and $\II$, then $\I$ becomes equivalent to $\III$, and $\II$ is equivalent to $\IV$. Hence one takes either $(\I,\III)$ or $(\II,\IV)$. Note that this reasoning is more general than appealing directly to the axiom (\ref{inde}), since this mathematical axiom does not specify the interpretation of the mixture model $\alpha p+(1-\alpha)r$; see Appendix \ref{inde3} for details. 

If experimental subjects are presented Allais' lotteries in the form (\ref{re1}, \ref{re2}), then majority of them behave according to EUT than for (\ref{al_1}, \ref{al_2}) \cite{baron}. Naturally, for the regret (\ref{r7}) the difference between (\ref{re1}, \ref{re2}) and (\ref{al_1}, \ref{al_2}) is absent. Hence these subjects did not use the regret theory in their decision making.

\section{Regret and stochastic dominance}
\label{regret_sto}

For lotteries (\ref{1}) with independent probabilities, a clear-cut definition of superiority is provided by the stochastic dominance $\succeq_{\rm sto}$ \cite{luce}. Recall its definition: we assume \footnote{This assumption of identical outcomes is not necessary, since the stochastic dominance can be formulated more generally. We do not focus on this general definition, since it is equivalent to the situation, when the outcomes are made the same by increasing their number via adding zero-probability events; cf.~the discussion before (\ref{comp}). } that $x_k=y_k$ in (\ref{1}) hold with 
\be
\label{ou}
x_{i} < x_{j}\quad {\rm for} \quad i<j. 
\ee
Now define \cite{luce}
\be
(x,p)\succeq_{\rm sto}(x,q)\quad {\rm iff} \quad
{\sum}_{i=1}^{k}p_{i}\leq {\sum}_{i=1}^{k} q_{i}\quad {\rm for}\quad k=1,..,n.
\label{sto}
\ee
Recall that the utility $u(x)$ in (\ref{r7}) is an increasing function of $x$. Stochastic dominance does not depend on a specific form of the utility $u(x)$ in (\ref{r7}) provided that it is an increasing function of $x$, as we assume. This is an advantage of stochastic dominance. Its weakness is that it clearly does not apply to all lotteries, i.e. the completeness axiom (\ref{comp}) is violated. Indeed, it is sufficient to violate (\ref{sto}) for one value of $k$, and this will make $\succeq_{\rm sto}$ inapplicable. A related weakness is that its applicability is not stable with respect to small variations of outcomes. To see this, assume that (\ref{ou}, \ref{sto}) hold and perturb $y_1=x_1\to y_1'<x_1$. Even a small variation of this type violates condition (\ref{sto}) for $k=1$.

Regret and stochastic dominance do not contradict each other, as the following proposition shows.

\vspace{0.2cm}
\underline{\textbf{Proposition 4.}} $(x,p)\succeq_{\rm sto}(x,q)$ implies $(x,p)\succeq_{\rm reg}(x,q)$ defined from (\ref{r7}). The proof is given in Appendix \ref{ap_b}.
\vspace{0.2cm}

Note that Proposition 4 does not require any specific feature of $g(x)$ apart of (\ref{u2}, \ref{uu2}). However, it does require independent probabilities for the lotteries, as implied by (\ref{r7}). Lotteries with independent probabilities have vast but still limited range of applications. Even within the framework of initially independent lotteries, one can envisage new dependent lotteries for which the regret is given via (\ref{r777}). For dependent lotteries the relation between regret and stochastic dominance is partially explained by the following proposition.

\vspace{0.2cm}
\underline{\textbf{Proposition 5.}} 
For the joint probability $P(x_i, x_j)$ given by (\ref{voche}), let us define the marginal probabilities $\{p_i\}_{i=1}^n$ and $\{q_j\}_{j=1}^n$, as well as deviation of $P(x_i, x_j)$ from $p_iq_j$:
\be
\label{bo}
&& p_i := {\sum}_{j=1}^{n}P(x_i, x_j),\quad q_j := {\sum}_{i=1}^{n}P(x_i, x_j),\\
\label{otark}
&& \theta_{i,j}:= P(x_i, x_j) - p_i q_j, \\
&& {\sum}_{i=1}^{n} \theta_{i,j} = {\sum}_{j=1}^{n}\theta_{i,j}=0, \qquad |\theta_{i,j}| \leq p_i q_j.
\label{gamb}
\ee
Then if $g(x)$ is super-additive on positive domain [see (\ref{al_8})] and if
\be
\theta_{i,j} \geq \theta_{j,i},\quad {\rm for}\quad i>j,
\label{buk}
\ee
one has that $(x,p)\succeq_{\rm sto}(x,q)$ defined via (\ref{ou}, \ref{sto}) leads to $(x,p)\succeq_{\rm reg}(x,q)$ in the sense of (\ref{r777}).

Thus the super-additivity of $g(x)$ plus condition (\ref{buk}) make the regret consistent with the stochastic dominance. The proof of Proposition 5 is given in Appendix \ref{bars}.

\section{Savage's omelet is solved via the regret theory}
\label{ap_a}

Eq.~(\ref{1}) with $\{p_k=q_k\}_{k=1}^n$ can refer to the to the decision model which assumes that at the moment of action-taking there is an uncertain state of nature (environment) ${\cal S}_k$ to be realized from $\{{\cal
S}_k\}_{k=1}^n$ with probabilities $\{p_k\}_{k=1}^n$, which are known to the decision maker \cite{luce,jeffrey}. 
${\cal S}_k$ are called states of nature, since their future realization is independent from the action taken, but an action $A$ ($B$) in a state ${\cal S}_k$ leads to consequences with monetary outcome $x_k$ ($y_k$) and utilities $u(x_k)$ ($u(y_k)$) \cite{luce,jeffrey}; cf.~(\ref{1}, \ref{r55}). 

The following classic decision problem is described in \cite{savage}: A decision maker has to finish making an omelet began by his wife, who has already broken into a bowl five good eggs. A sixth unbroken egg is lying on the table, and it must be either used in making the omelet, or discarded. There are two states of the nature: good (the sixth egg is good) and rotten (the sixth egg is rotten), which do not depend on the actions $A_1$, $A_2$ and $A_3$ of the decision maker. 

$A_1$: break the sixth egg into the bowl. 

$A_2$: discard the sixth egg. 

$A_3$: break the sixth egg into a saucer; add it to the five eggs if it is good, discard it if it is rotten. 

The consequences of the acts can be written as lotteries: 
\be
A_1= \begin{pmatrix}
u_{-5} & u_6\\
p & 1-p
\end{pmatrix}, \qquad
A_2= \begin{pmatrix}
u_5 & u_5+z \\
p & 1-p
\end{pmatrix}, \qquad
A_3= \begin{pmatrix}
u_5+w & u_6+w \\
p & 1-p
\end{pmatrix},
\label{aa1}
\ee
where $p$ ($1-p$) is the objective probability for the sixth egg to be rotten (good), $u_6$ ($u_5$) is the utility of the six-egg (five-egg) omelet, $u_{-5}<0$ is the utility of five spoiled eggs and no omelet whatsoever, $w<0$ is the utility of washing the saucer, and $z<0$ is the utility of the good egg being lost.
\footnote{The concrete utilities of washing the saucer may differ depending on the state of the sixth egg. We, however, neglect this difference. Also, for simplicity $w$ was simply added to $u_5$ and $u_6$.}

Now looking at the consequences of $A_2$, we see that|in contrast to $A_1$ and $A_3$|acting $A_2$ does not resolve the uncertain state of nature: once the egg is discarded, the decision maker will not know (without additional actions), whether it was rotten or good. Put differently, utility $z$ is not obtained after acting $A_2$, and cannot be obtained without additional actions. Calculating the expecting utility of $A_2$ in the usual way as $p u_5+(1-p)(u_5+z)$ does not apply, because it disregards this aspect $A_2$. It is  natural to take the expected utility as $p u_5+(1-p)u_5=u_5$ (i.e. once $z$ is not obtained, it is not included), but then comparing with expected utilities of $A_1$ and $A_3$, we see that the parameter $z$ will appear nowhere. Hence, we suggest that the expected utility does not apply to comparing $A_2$ with the other two actions. 

Employing in (\ref{aa1}) the reasoning of regret [cf.~(\ref{r1}, \ref{r44}, \ref{r55})]  does take into account the difference between $A_2$ and the other two actions. Let us for example calculate the regret about not taking $A_1$ once $A_2$ has been taken:
\be
R(A_1,A_2)=pf(u_{-5}-u_5)+(1-p)f(u_6-u_5).
\label{aa3}
\ee
This expression does not contain $z$, since the uncertain state of nature was not resolved after acting $A_2$, i.e. after acting $A_2$ the obtained utility is $u_5$. 

On the other hand, acting $A_1$ resolves the uncertainty about the state of nature. Hence the regret of not taking $A_2$, once $A_1$ was acted reads:
\be
R(A_2,A_1)=pf(u_5-u_{-5})+(1-p)f(u_5+z-u_6),
\label{aa4}
\ee
i.e. once $A_1$ is taken and the egg is rotten (good), then the decision maker already knows that if $A_2$ would be taken, then the egg will turn out rotten (good). It is seen that (\ref{aa4}) contains $z$ (the utility of discarding a good egg), while (\ref{aa3}) does not. Now 
\be
A_1\succeq_{\rm reg}A_2 \quad {\rm iff} \quad R(A_2,A_1)\leq R(A_1,A_2),
\label{owl}
\ee
where $R(A_2,A_1)- R(A_1,A_2)$ does feel the parameter $z$. As an example of (\ref{owl}) consider $f(x)=x$ [cf. the discussion after (\ref{rev})]:
\be
p(u_5-u_{-5})<(1-p)(u_6-u_5-\frac{z}{2}).
\label{owl2}
\ee
where we recall that $u_6>u_5>u_{-5}$ and $z<0$. We can naturally assume $u_{5}-u_{-5}>u_6-u_5>0$ under which 
(\ref{owl2}) is non-trivial even for $p=1/2$. Note that the formal application of the expected utility will claim that $A_1$ is preferred over $A_2$ for $p(u_5-u_{-5})<(1-p)(u_6-u_5-z)$, which is clearly different from (\ref{owl2}). This is not just a different outcome; rather, the expected utility does not apply.

\section{Summary}

This paper studied regret functionals over the utility differences of two probabilistic lotteries; see section \ref{regret_basic}. There are various types of lotteries, from independent to fully dependent that refer to the same state of nature. The regret functional compares the lotteries counter-factually taking notice of their probabilities. For particular cases, the regret reverts to the expected utility. More generally, it does not satisfy the independence (from irrelevant alternatives) axiom of the expected utility, also known as the sure thing principle. This not satisfying is by itself non-trivial and is explored in Appendix \ref{inde3}. In contrast to the expected utility, the regret is also generally not invariant with respect to multiplying the utility by a positive number. It is also due to these two differences compared to the expected utility that the regret is efficient for explaining and resolving Allais's paradox; see section \ref{alais_solution}. The resolution demands a non-trivial features of the regret functional, {\it viz.} its super-additivity, which does make an intuitive sense. We show that the regret functional can be chosen such that the regret-ordering holds transitivity. In particular, Allais's paradox can be resolved via a transitive regret, and this resolution provides a consistent account of changes in monetary outcomes.

We devoted a special attention to relations between (the first-order) stochastic dominance and the regret-preference; see section \ref{regret_sto}. The former ordering is normatively appealing, but it is incomplete, since not every two lotteries can be compared with each other. We show that for independent lotteries the stochastic dominance implies the regret-preference. For dependent lotteries the relations between the two are more complex. Here we proposed a sufficient condition for the implication stochastic dominance $\to$ regret-preference, which, interestingly is also based on the super-additivity of the regret; see Proposition 5.

Finally, we show in section \ref{ap_a} how the considered regret theory can be useful in those situations, where actions of the decision maker do not resolve the uncertain situation. The expected utility theory does not apply to such a situation in the sense that there is an important information about the lotteries that it simply discards. 
In the regret, this information is employed, since the regret compares the unresolved uncertainty with the resolved uncertainty.

Our results show that though the concept of regret was initially deduced from certain emotional features of decision makers, it does have many features one intuitively expects from rationality. Hence we envisage its further applications in e.g. reinforcement learning.

\section*{Acknowledgements}

This work was supported by State Science Committee of Armenia, grants No. 21AG-1C038. We thank Andranik Khachatryan for useful remarks and for participating in initial stages of this work. 

\bibliographystyle{ieeetr}
\bibliography{Allais_regret}

\appendix

\section{Proof of Proposition 4}.
\label{ap_b}

\noindent We are to show that
\begin{equation}
\label{b-1}
    \sum_{i=1,j=1}^{n}p_{i}q_{j}g(u(x_j)-u(x_i))\leq 0.
\end{equation}
 It can be written in the following form
\begin{equation}
    \sum_{i=1}^{n}\sum_{j=1}^{i}(p_{i}q_{j}-p_{j}q_{i})g(u(x_j)-u(x_i)).
    \label{b-3}
\end{equation}
Now $g$'s in (\ref{b-3}) are strictly negative (recall (\ref{ou}) and monotonicity of $g(x)$ and $u(x)$).

\noindent We show that the following holds
\begin{equation}\label{b_main}
\begin{split}
     \sum_{i=1}^{n}\sum_{j=1}^{i} p_{i}q_{j}g(u(x_j)-u(x_i)) & \leq \sum_{i=1}^{n}\sum_{j=1}^{i} p_{i}p_{j}g(u(x_j)-u(x_i)) \\ 
     & \leq  \sum_{i=1}^{n}\sum_{j=1}^{i} p_{j}q_{i}g(u(x_j)-u(x_i)),
\end{split}
\end{equation}
 from which the (\ref{b-3}) will follow.

\noindent To do it, we make use of the following lemma.

\noindent \textbf{Lemma}. Given sequences $p_{i}$, $q_{i}$ as above, then for any  increasing sequence of negative numbers $g_i$, it is true that
\begin{equation}
    \sum_{i=1}^{k}p_{i}g_{i} \geq \sum_{i=1}^{k}q_{i}g_{i}, \quad {\rm for} \quad k=1,...,n.
\end{equation}
 A simple induction will help.
 Obviously $p_{1}g_{1} \geq q_{1}g_{1}$. Supposing it is true for some $k$ we will have:
\begin{equation}
    \sum_{i=1}^{k}p_{i}g_{i} \geq \sum_{i=1}^{k}q_{i}g_{i}
\end{equation}
Let us now subtract $g_{k+1}$ from each component $g_{i}$. (As we are speaking about any increasing sequence, we haven't specified any concrete $g$-s yet. 
So, having that all components are less $g_{k+1}$,  subtract, without change of sign in inequality)

\noindent We will have the following:
\begin{equation}\label{bfirst}
    \sum_{i=1}^{k}p_{i}(g_{i}-g_{k+1}) \geq \sum_{i=1}^{k}q_{i}(g_{i}-g_{k+1})
\end{equation}
 Now by (\ref{sto})
\begin{equation}
\sum_{i=1}^{k+1}p_{i} \leq \sum_{i=1}^{k+1}q_{i}
\end{equation}
multiplying by $g_{k+1}$, we get
\begin{equation}\label{bsecond}
\sum_{i=1}^{k+1}p_{i}g_{k+1} \geq \sum_{i=1}^{k+1}q_{i}g_{k+1}.
\end{equation}
Summing up (\ref{bfirst}) and (\ref{bsecond}), the desired result is obtained:
\begin{equation}
\sum_{i=1}^{k+1}p_{i}g_{i} \geq \sum_{i=1}^{k+1}q_{i}g_{i}.
\end{equation}
Using the lemma we go back and consider first part of (\ref{b_main})
\begin{equation}
     \sum_{i=1}^{n}\sum_{j=1}^{i} p_{i}q_{j}g(u(x_j)-u(x_i))  \leq \sum_{i=1}^{n}\sum_{j=1}^{i} p_{i}p_{j}g(u(x_j)-u(x_i)).
\end{equation}
Obviously 
\begin{equation}
\begin{split}
    \sum_{i=1}^{n}\sum_{j=1}^{i} p_{i}q_{j}g(u(x_j)-u(x_i)) & = \sum_{i=1}^{n} p_{i}\sum_{j=1}^{i}q_{j}g(u(x_j)-u(x_i)) \\ & \leq
 \sum_{i=1}^{n} p_{i}\sum_{j=1}^{i}p_{j}g(u(x_j)-u(x_i)),
 \end{split}
\end{equation}
The last part is implied by lemma.

\noindent Returning to the second part of (\ref{b_main}):
\begin{equation}
    \sum_{i=1}^{n}\sum_{j=1}^{i} p_{j}q_{i}g(u(x_j)-u(x_i))   \geq \sum_{i=1}^{n}\sum_{j=1}^{i} p_{i}p_{j}g(u(x_j)-u(x_i)).
\end{equation}
and changing the order of summation, we get
\begin{equation}
\begin{split}
 \sum_{i=1}^{n}& \sum_{j=1}^{i} p_{j}q_{i}g(u(x_j)-u(x_i))= 
\sum_{j=1}^{n}\sum_{i=j}^{n} p_{j}q_{i}g(u(x_j)-u(x_i)) \\
& =\sum_{j=1}^{n}p_{j}\sum_{i=j}^{n} q_{i}g(u(x_j)-u(x_i)) \geq
\sum_{j=1}^{n}p_{j}\sum_{i=j}^{n} p_{i}g(u(x_j)-u(x_i))  \\
&=\sum_{i=1}^{n}\sum_{j=1}^{i} p_{i}p_{j}g(u(x_j)-u(x_i)),
\end{split}
\end{equation}
where the inequality is the inverse of the one used in lemma. 
The proof is complete.

\section{Regret and the independence axiom (\ref{inde})}
\label{inde3}

We already emphasized around (\ref{re1}, \ref{re2}) that regret preference $\succeq_{\rm reg}$ defined in (\ref{r7}) must violate the independence axiom for solving Allais' paradox. Now we provide clarifications regarding the form (\ref{inde}) of this axiom. We note that (\ref{cont}, \ref{inde}) do not define how precisely the mixing of the two probabilities with weights $\alpha$ is implemented. Below we discuss three interesting possibilities for implementing the set-up of (\ref{inde}). So we are given three lotteries [cf.~(\ref{1})]
\begin{gather}
\label{s1}
(x,p)=
\begin{pmatrix}
x_{1} & x_{2} & ... & x_{n}\\
p_{1} & p_{2} & ... & p_{n}
\end{pmatrix},\qquad
(y,q)=
\begin{pmatrix}
y_{1} & y_{2} & ... & y_{n}\\
q_{1} & q_{2} & ... & q_{n}
\end{pmatrix},
\qquad
(z,r)=
\begin{pmatrix}
z_{1} & z_{2} & ... & z_{n}\\
r_{1} & r_{2} & ... & r_{n}
\end{pmatrix},\\
{\sum}_{k=1}^np_k={\sum}_{k=1}^nq_k={\sum}_{k=1}^nr_k=1,
\end{gather}
where $(p_1,...,p_n)$, $(q_1,...,q_n)$, and $(r_1,...,r_n)$ are (resp.) the probabilities of monetary outcomes
$(x_1,...,x_n)$ and $(y_1,...,y_n)$, $(z_1,...,z_n)$ within each lottery.

{\bf 1.} Here one chooses between two composite lotteries $A=\{(1-\alpha)(x,p)+\alpha (z,r)\}$ and $B=\{(1-\alpha)(y,q)+\alpha (z,r)\}$. If $A$ is taken, then a binary random variable $S_A$ is realized that takes values $S_A=0$ and $S_A=1$ with probabilities $1-\alpha$ and $\alpha$, respectively. For $S_A=0$ or $S_A=1$ one faces lottery $(x,p)$ or $(z,r)$, respectively. If $B$ is taken, then a binary random variable $S_B$ (independent from $S_A$) is realized that takes values $S_B=0$ and $S_B=1$ with probabilities $1-\alpha$ and $\alpha$, respectively. For $S_B=0$ or $S_B=1$ one faces lottery $(y,q)$ or $(z,r)$, respectively. Let us now assume that $(x,p)$ is independent from $(y,q)$, but the lottery $(z,r)$ in both options is the same. Using definition of regret (\ref{r7}), we end up with the following preference relation:
\be
&&A\succeq_{\rm reg, 1} B\quad {\rm iff}\quad
(1-\alpha)^2{\sum}_{i,j=1}^n g(u(y_j)-u(x_i))p_iq_j 
+ (1-\alpha)\alpha{\sum}_{i,j=1}^n g(u(z_j)-u(x_i))p_ir_j\nonumber\\
&&+(1-\alpha)\alpha {\sum}_{i,j=1}^n g(u(y_j)-u(z_i))q_jr_i \leq 0.
\label{ar777}
\ee
Note that (\ref{ar777}) does not contain terms with $g(u(z_j)-u(z_i))$, because the decision maker does not expect to find different outcomes $z_j$ and $z_i$ within options $A$ and $B$.

{\bf 2.} Now we have the situation of {\bf 1}, but $S=S_A=S_B$; e.g. one can assume that $S$ is realized beforehand, but the result is not known to the decision maker at the time of decision making. Now the regret is different [cf.~(\ref{ar777})]:
\be
A\succeq_{\rm reg, 2} B\quad {\rm iff}\quad
(1-\alpha)^2{\sum}_{i,j=1}^n g(u(y_j)-u(x_i))p_iq_j\leq 0,
\label{ar778}
\ee
where $(z,r)$ does not enter to regret comparison (\ref{ar778}), which is formally consistent with axiom (\ref{inde}).

{\bf 3.} We have the situation of {\bf 1}, but $(z,r)$ in option $A$ and $(z,r)$ in option $B$ are two different lotteries with independent probabilities:
\be
&&A\succeq_{\rm reg, 3} B\quad {\rm iff}\quad
(1-\alpha)^2{\sum}_{i,j=1}^n g(u(y_j)-u(x_i))p_iq_j 
+ (1-\alpha)\alpha{\sum}_{i,j=1}^n g(u(z_j)-u(x_i))p_ir_j \nonumber\\
&&+(1-\alpha)\alpha {\sum}_{i,j=1}^n g(u(y_j)-u(z_i))q_jr_i 
+\alpha^2 {\sum}_{i,j=1}^n g(u(z_j)-u(z_i))r_jr_i \leq 0.
\label{ar779}
\ee
The standard interpretation of the independence axiom within the expected utility theory hints at {\bf 3}. We however emphasized that this situation is not unique. Note that all possibilities (\ref{ar777}, \ref{ar778}, \ref{ar779}) agree with the expected utility theory, where $g(x)=x$. 

\section{Proof of Proposition 5}
\label{bars}

It is known that $n \times n$ matrices $\theta_{i,j}$ from (\ref{otark}) form vector space of $(n-1)\times (n-1)$ dimension. So for example any such $3 \times 3$ matrix can be rewritten
\begin{equation}
    \begin{pmatrix}
    \theta_{1,1} & \theta_{1,2} & \theta_{1,3} \\
    \theta_{2,1} & \theta_{2,2} & \theta_{2,3} \\
    \theta_{3,1} & \theta_{3,2} & \theta_{3,3} 
    \end{pmatrix} = 
    \theta_{1,1}  \begin{pmatrix}
    1 & 0 & -1 \\
    0 & 0 & 0 \\
    -1 & 0 & 1 
    \end{pmatrix} + \theta_{1,2}  \begin{pmatrix}
    0 & 1 & -1 \\
    0 & 0 & 0 \\
    0 & -1 & 1 
    \end{pmatrix} +  \theta_{2,1}  \begin{pmatrix}
    0 & 0 & 0\\
    1 & 0 & -1 \\
    -1 & 0 & 1 
    \end{pmatrix} +  \theta_{2,2}  \begin{pmatrix}
    0 & 0 & 0 \\
    0 & 1 & -1 \\
    0 & 1 & 1 
    \end{pmatrix}
\end{equation}
Denoting the basis matrices by $M_{i,j}$ we have that any matrix $\Theta$ of thetas can be rewritten as 
\be
\Theta = \sum_{i,j = 1} ^{n-1} \theta_{i,j} M_{i,j}, 
\ee
where $M_{i,j}$ is the matrix whose $(i,j)$-th and $(n,n)$-th elements are $1$, the $(i, n)$-th and $(n, j)$-th elements are $-1$.

Let us compute the regret in this case
\begin{equation}
    R = \sum_{i=1}^{n}\sum_{j=1}^{n} P(x_i, x_j) g(u(x_j)-u(x_i)) = \sum_{i=1}^{n}\sum_{j=1}^{n} p_i q_j g(u(x_j)-u(x_i)) +  \sum_{i=1}^{n}\sum_{j=1}^{n} \theta_{i,j} g(u(x_j)-u(x_i)) 
\end{equation}
We already know that the first term is negative ((\ref{b-1}) and Proposition 4). So it remains to show that second part is also negative. 
Denoting by $G$ the matrix whose elements are $G(i,j) = g(u(x_j)-u(x_i))$, we can rewrite 
\begin{equation}
    \sum_{i=1}^{n}\sum_{j=1}^{n} \theta_{i,j} g(u(x_j)-u(x_i)) = ||\Theta \odot G||,
\end{equation}
Where under $||.||$ we understand sum of all elements and $\odot$ is Hadamard's (element-wise) product.

\noindent Note that
\begin{equation}
    \sum_{i=1}^{n}\sum_{j=1}^{n} \theta_{i,j} g(u(x_j)-u(x_i)) = ||\Theta \odot G||=\sum_{i=1}^{n-1} \sum_{j=1}^{n-1}\theta_{i,j} ||M_{i,j} \odot G||.
\end{equation}
We have
\begin{equation}
     ||M_{i,j} \odot G|| = g(u(x_i) - u(x_j)) - g(u(x_n) - u(x_j)) - g(u(x_i) - u(x_n)).
\end{equation}
Now consider
\begin{equation}
\begin{split}
   & \theta_{i,j} ||M_{i,j} \odot G||  +\theta_{j,i} ||M_{i,j} \odot G|| \\ &  = \theta_{i,j} ( g(u(x_i) - u(x_j)) - g(u(x_n) - u(x_j)) - g(u(x_i) - u(x_n))) \\ & + \theta_{j,i} ( g(u(x_j) - u(x_i)) - g(u(x_n) - u(x_i)) - g(u(x_j) - u(x_n))) \\ &  = (\theta_{i,j} - \theta_{j,i}) (g(u(x_i)-u(x_j))+g(u(x_n)-u(x_i))- g(u(x_n)-u(x_j))
\end{split}
\end{equation}
Note that while $i>j$ we have by super-additivity that term in second parenthesis is negative. So toghether with $\theta_{i,j} \geq \theta_{j,i}$ we conclude that
\begin{equation}
    \theta_{i,j} ||M_{i,j} \odot G||  +\theta_{j,i} ||M_{i,j} \odot G|| \leq 0
\end{equation}

Rewriting

\begin{equation}
    \sum_{i=1}^{n}\sum_{j=1}^{n} \theta_{i,j} g(u(x_j)-u(x_i)) = ||\Theta \odot G||=\sum_{i=1}^{n-1} \sum_{j=1; j \neq i}^{n-1}\theta_{i,j} ||M_{i,j} \odot G|| +\sum_{i=1}^{n-1} \theta_{i,i} ||M_{i,i} \odot G||.  
    \label{oko}
\end{equation}
The second sum in (\ref{oko}) is obviously 0, as $g(x)$ is antisymmetric, and $M_{i,i}$-s are symmetric matrices. 

\noindent So 
\begin{equation}
    \sum_{i=1}^{n}\sum_{j=1}^{n} \theta_{i,j} g(u(x_j)-u(x_i)) =\sum_{i=1}^{n-1} \sum_{j=1; j \neq i}^{n-1}\theta_{i,j} ||M_{i,j} \odot G|| \leq 0 .
\end{equation}

\comment{
\section{Prospect theory}

Within the prospect theory the value of each lottery reads:
\begin{equation}
\label{pro}
    V(x) = {\sum}_{i=1}^n \pi (p_i) u(x_{i}).
\end{equation}
In the original prospect theory, the utility function $u(x;x_0)$ is concave (convex) for $x>x_0$ ($x<x_0$). Here the parameter $x_0$ is related with the initial wealth. positive and convex when it is negative. In (\ref{pro}), $\pi(.)$ is probability weighting function.
The idea behind it is that people facing choices between lotteries tend to re-estimate the probabilities exaggerating the small ones, and decreasing the big ones.
However, this understanding of $\pi(.)$ was later revised, because people subjectively tend to overestimate the small probabilities of "bad" outcomes, and to reduce high probabilities of "good" ones.
This brought to alternative forms of prospect theory, where $\pi$ is applied to cumulative probabilities \cite{Cumul_Prospect}. 

The prospect theory can explain Allais paradox, but the properties of utility and probability weighting functions were not devised from anything other then experimental argumentation. 
}


\end{document}